\documentclass[conference]{IEEEtran}
\usepackage{cite, epsfig, mathptmx, times, amssymb, amsmath, color}
\usepackage{subcaption}

\usepackage{graphicx,booktabs,soul,color,graphics,epsfig,mathptmx,moreverb,ifpdf,cancel,amsthm,framed,dblfloatfix,array}
\usepackage{gensymb,graphicx,url, multirow,kantlipsum}
\usepackage{fancyhdr}
\usepackage{tabularx,caption}
\usepackage{mhchem}
\usepackage{siunitx}
\usepackage{multirow}
\usepackage{tabularx}
\newcolumntype{Z}{>{\raggedright}X}

\usepackage{lipsum}  

\newcolumntype{F}[1]{%
    >{\raggedright\arraybackslash\hspace{0pt}}p{#1}}%
\newcolumntype{T}[1]{%
    >{\centering\arraybackslash\hspace{0pt}}p{#1}}%

\makeatother

\DeclareRobustCommand*{\IEEEauthorrefmark}[1]{%
	\raisebox{0pt}[0pt][0pt]{\textsuperscript{\footnotesize #1}}%
}

\begin{document}

\title{Understanding long-term energy use in off-grid solar home systems in sub-Saharan Africa}

\author{\IEEEauthorblockN{\IEEEauthorblockN{
			R. Perriment\IEEEauthorrefmark{1}, 
            V. Mergulhao\IEEEauthorrefmark{2},
            V. Kumtepeli\IEEEauthorrefmark{1},
            P. Parikh\IEEEauthorrefmark{2},
            M.D. McCulloch\IEEEauthorrefmark{1},
			D. A. Howey\IEEEauthorrefmark{1}
		}
		\IEEEauthorblockA{
			\IEEEauthorrefmark{1}Department of Engineering Science, University of Oxford, Oxford, OX1 3PJ, United Kingdom \\
            \IEEEauthorrefmark{2}UCL Engineering for International Development Centre, The Bartlett School of Sustainable Construction \\
            University College London, Gower Street, London WC1E 6BT, United Kingdom
			}}}

\maketitle

\begin{abstract}
Solar home systems provide low-cost electricity access for rural off-grid communities. As access to them increases, more long-term data becomes available on how these systems are used throughout their lifetime. This work analyses a dataset of 1,000 systems across sub-Saharan Africa. Dynamic time warping clustering was applied to the load demand data from the systems, identifying five distinct archetypal daily load profiles and their occurrence across the dataset. Temporal analysis reveals a general decline in daily energy consumption over time, with 77\% of households reducing their usage compared to the start of ownership. On average, there is a 33\% decrease in daily consumption by the end of the second year compared to the peak demand, which occurs on the 96th day. Combining the load demand analysis with payment data shows that this decrease in energy consumption is observed even in households that are not experiencing economic hardship, indicating there are reasons beyond financial constraints for decreasing energy use once energy access is obtained. 
\end{abstract}


\section{Introduction}

Sustainable Development Goal 7 (SDG7) highlights the need for affordable, reliable, and sustainable energy access for all. Meeting this pressing challenge requires a dual focus on transitioning to renewable energy sources while simultaneously ensuring energy access for the 759 million people without electricity worldwide, 75\% of whom live in sub-Saharan Africa \cite{GoalAffairs}. Abundant solar resources and dramatically decreasing costs of photovoltaic panels make solar-based technology solutions a prime contender for sustainable electrification of sub-Saharan Africa.

While there has been significant progress towards universal electrification, the rate of global electrification has slowed in recent years due to the increasing difficulty of providing access to the remaining un-electrified population, who often live in remote, isolated areas where grid extension is not feasible.  \cite{Almeshqab2019LessonsAspects}. The popularity of mini-grids to serve these regions is growing, but they require large upfront investments, which in turn require guaranteed sufficient demand \cite{Narayan2020TheChallenges}. On the other hand, solar home systems (SHSs) are a viable alternative with lower upfront costs for providing Tier 1 and 2 energy access. (Tier 1 energy access includes electricity for lighting, radio and phone charging with a minimum of 12 Wh/day, while Tier 2 access includes all Tier 1 functions plus televisions and fans with a minimum of 200 Wh/day.) The simplicity and lower upfront costs of SHSs make them an attractive choice---across Africa, 74\% of off-grid energy investments between 2010 and 2020 were in SHSs \cite{Irena2022RENEWABLEREGIONS}.

While the roll-out of smart meters in high-income countries (HICs) has led to an influx of data on electricity consumption, there is a lack of data on electricity consumption in low- and lower-middle-income countries (LMICs). This is particularly true for newly electrified households where there was no previous access to electricity. In this paper we analyse a large SHS dataset from the Africa-based energy supplier BBOXX, applying clustering techniques to understand typical load profiles in newly electrified households. The payment structure used by BBOXX for their SHS is also very different to the standard price-per-kWh format used in HIC. Instead, a pay-as-you-go (PAYG) system is used. The amount each household pays is fixed per day of using the system, and based on the number of appliances the system powers. The PAYG approach allows users to top-up energy when they desire, in their local currency, and enables them to stop paying on certain days if they wish. However, when payment stops, electricity usage is also stopped.

Increasing the availability of rural off-grid household load data is important for many stakeholders, from system designers to aid sizing and design choices, to local governments to aid policy decisions, to researchers trying to improve off-grid systems and their use. We analyse the energy consumption of over 1,000 households with SHS across sub-Saharan Africa. Initially, we cluster the time series data to obtain representative load profiles. We then analyse the changes in household energy use over time by looking at shifts from cluster to cluster, and compare this with payment data to better understand the drivers for energy behaviour.

\section{Literature Review}

Understanding the electrical load requirements of rural off-grid households is essential to understand individuals needs and ensure appropriate design and maintenance choices are made. Before access to big field datasets was possible, the common method of estimating load demand was consumer use surveys, where households were asked questions on anticipated appliance ownership and times of use \cite{Pandyaswargo2020EstimatingLaos, Sandwell2016AnalysisPradesh, Mandelli2016NovelAreas, Boait2017ESCoBox:World, Boait2015EstimationCountries}. However, surveys have since been proven to be an inaccurate indicator of load demand, both in terms of magnitude and time of use, with errors up to four times greater than actual demand data \cite{Blodgett2017AccuracyUserconsumption, Hartvigsson2018ComparisonData, Allee2021PredictingModels,Gelchu2023ComparisonEthiopia}. 

A data-driven approach is established as more accurate in literature, however, the lack of available field data is a challenge \cite{Bisaga2017ScalableSolar}. In some cases, a limited amount of load demand data is available and this can be used alongside other techniques. Allee et al.\ \cite{Allee2021PredictingModels} combine a data-driven approach with survey results to predict the load demand of 1,378 Tanzanian mini-grid customers, demonstrating the use of surveys can be beneficial when supplemented by measured data. Few et al.\ \cite{Few2022ElectricityDesign} combine mini-grid load demand data with information on climatic conditions and household level characteristics, such as appliance ownership, to simulate household level load profiles. Both cases show that additional information, beyond stand-alone load demand data, further increases load demand estimation capabilities. Yet, reliable load demand data is still the essential starting block.

Greater granularity of the load demand data enables useful insights for system design and maintenance, but sometimes aggregated data is all that is available. Data can be aggregated on a household level (i.e.\ for multiple households) \cite{Bhattacharyya2021AAsia} or a temporal level (i.e.\ multiple households for a single time period) \cite{DenHeeten2017UnderstandingElectrification}, but in both instances, the results are diminished compared to using raw granular data, and this can negatively impact design and maintenance choices \cite{Jurasz2022OnSystems}.

Where ample data is available, clustering can be a useful tool for visualisation and understanding. In the case of load demand clustering, k-centered methods are a common choice due to simple implementation \cite{Yilmaz2019ComparisonManagement}. Some authors have made efforts to compare the effectiveness of different clustering methods \cite{Rajabi2020ASegmentation, Wang2015LoadReview, McLoughlin2015AData}, but different results are found in each case. This highlights the importance of having a clear goal for the clustering process, since this ensures that the algorithms and distance metrics output logical and relevant clusters \cite{VonLuxburg2012Clustering:Art, Hennig2015WhatClusters}. 

Dynamic time warping (DTW) is an elastic distance metric for comparing time-series data which has a one-to-many mapping and is robust to shifts in time \cite{Sakoe1978DynamicRecognition}. This allows a shape-based comparison, as proposed in \cite{Lin2019ClusteringApplications}, without being dependent on dimensionality reduction. The benefits of using DTW over Euclidean distance in load profile clustering are beginning to be recognised in literature \cite{Aghabozorgi2015Time-seriesReview, Ausmus2020ImprovingWarping, Begum2015AcceleratingStrategy, Ratanamahatana2005ThreeMining}. However, due to its quadratic computational complexity, it is not commonly implemented \cite{Rajabi2020ASegmentation}. Teeraratkul et al.\ \cite{Teeraratkul2016CondensedDemand} and Ausmus et al.\ \cite{Ausmus2020ImprovingWarping} clustered household load profiles using DTW with k-medoids and k-means respectively. Both found significant improvement compared to using Euclidean distance, and k-medoids has the advantage of using a time series from the dataset as the representative cluster centre, which can improve interpretability. However, use of k-centered methods in the cluster assignment stage makes the algorithm vulnerable to getting stuck in local optima because the iterative method to solve the k-medoids and k-means problem does not guarantee global optimality. Kumtepeli et al.\ \cite{Kumtepeli2024DTW-C++:Data} instead formulated the k-medoids problem as a mixed integer programming (MIP) problem, allowing global optimality within the cluster assignment while using DTW for the distance metric. MIP is beneficial for alleviating the risk of getting stuck in local optima, however using MIP on very large datasets (with respect to number of time series) is computationally expensive and unfeasible. 

Clustering load profiles from LMIC is much less common than load data for HIC in literature because less data is available. Williams et al.\ \cite{Williams2017LoadMicrogrids} characterised load profiles from eleven mini-grids across Africa finding trends in consumption on daily and monthly time scales. However, no clustering was done, and household level data was not made available, so comprehension of household trends is not possible. Lorenzoni et al.\ \cite{Lorenzoni2020ClassificationApproach} conducted hierarchical clustering on the load demand data of 61 mini-grids in LMICs, finding clusters of aggregated household data. The key differentiation between the clusters was the peaks in load demand. The authors also found that as the systems aged, and the energy connection tier increases, the use profile became flatter.
Lukuyu et al.\ \cite{Lukuyu2023PurchasingAfrica} used k-means to cluster daily load demand profiles of SHS customers in East Africa over a year. However, they used a monthly mean to represent the daily load profile for each household, which loses the granularity of day-to-day variability and gives smoothed profiles for the results.

In addition to understanding day-to-day energy use, it is important to consider long-term behaviour, particularly in the context of energy access for previously unelectrified households. In these situations, behaviour is more unpredictable \cite{Riva2018Long-termPerspective,Muhumuza2018EnergyCountries}. Therefore, long-term load demand analysis is imperative to gain insights into how systems deployed in LMICs are used. %
It is commonly thought that energy demands will increase as energy access is obtained \cite{Riva2018Long-termPerspective,Muhumuza2018EnergyCountries,Opiyo2020HowCommunities}, however, this conjecture needs to be supported by real-world data \cite{Riva2019ModellingPlanning}. Bisaga et al.\ \cite{Bisaga2018ToLens} compared energy use of households across Rwanda, splitting them into groups depending on the length of time they had been using systems. They found that households that had systems for over a year generally used less energy each day than those who had obtained systems within the last 6 months, despite the more established households having generally more appliances. Kizilcec et al.\ \cite{Kizilcec2022ForecastingAccess} also conducted work on Rwandan SHS energy use, finding a reduction in electricity demand over the first year of use for the majority of households in their dataset. The decrease in electricity consumption was even more pronounced in households that owned a TV.

Beyond the first year of ownership, Dominguez et al.\ \cite{Dominguez2021UnderstandingHouseholds} found a general increase in monthly electricity expenditure after initial adoption, along with an increase in appliance ownership, but this plateaued at the 2-3 year mark. Conversely, \cite{Masselus202410Rwanda} found that rural households in Rwanda with grid connections decreased their electricity consumption over the first ten years of connection. This analysis includes monthly electrical consumption data from 147,074 rural households, focusing on 174 households with survey data. The surveys found that household appliances largely consisted of lighting and entertainment devices, but economically productive appliances were a rarity. Louie et al.\ \cite{Louie2023DailyNation} analysed the load demand of off-grid households in the Navajo Nation over 2 years and found a large variation both between and within households, the latter being impacted by seasonal changes. Additionally, energy use in the second year was on average 10\% lower than in the first year, highlighting longer-term temporal load shifts. 

The overall picture for long-term temporal changes in energy use remains unclear---increased consumption is anticipated but there are also datasets indicating the opposite. Further information is required to understand underlying causes. The barrier to rural electrification is often economic \cite{Blimpo2020WhyAfrica, Kizilcec2020SolarAfrica}, but there is a research gap in understanding the link between diverse PAYG payment data \cite{Mergulhao2023HowKenya} and energy use. Existing studies establish a link between energy use and payment, but focus mainly on payment behaviour impacts \cite{Guajardo2019HowEconomies} and very coarse-grained energy consumption data \cite{Lukuyu2023PurchasingAfrica}.

This work makes several contributions: firstly, a large-scale analysis of daily SHS load-demand profiles is undertaken, elucidating patterns in real-world data; secondly, progression of energy use over several years is considered, rather than observing a single snapshot in time; finally, payment and energy data are combined, enabling insight into the role of economic factors. This research also provides access to the measured data for others to use, supporting the collective efforts of researchers and industry to achieve SDG7.

\section{Energy use dataset and analysis}

\begin{figure}
    \centering
    \includegraphics[width=1\linewidth]{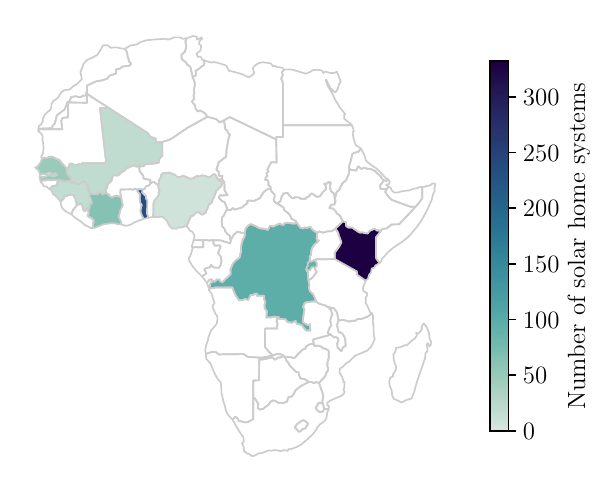}
    \caption{Geographical distribution of the downselected solar home systems used in the clustering process. The countries coloured white hosted no systems in our dataset.}
    \label{fig:country2}
\end{figure}

The dataset analysed in this paper was provided by BBOXX, and consists of time series from 1,000 solar home systems across Africa, each with a \SI{50}{W} solar panel, a \SI{12}{V}, \SI{17}{Ah} lead-acid battery, and various loads, including lighting, phone charging, fans, radios and TVs. These systems were downselected from the wider BBOXX database using stratified sampling based on system lifetime so that a range of systems were represented (including new adopters and more established users). To enable longer-term analysis, all SHSs younger than 1 year were excluded, and the oldest system in the dataset is 1,230 days old. The geographical distribution of systems in sub-Saharan Africa is shown in Fig.\ \ref{fig:country2}, with most located in Kenya, Togo, Democratic Republic of the Congo and Rwanda. The time for each system (normally stored as UTC) was corrected to local time in each country to ensure comparable solar and usage conditions.

The time, output current, and output voltage were used for load-demand analysis, with non-uniform sampling, i.e., measurements were recorded whenever a change above a certain threshold was detected. This was at second resolution, with a 10-minute backstop recording taken if no change was detected within that time. The following method was used to convert these measurements to hourly load demands:
\begin{enumerate}
    \item Calculate instantaneous power $P(t)$ at each time $t$ from measured voltage $V(t)$ and load current $I_\text{out}(t)$,
    \begin{equation}
        P(t)=V(t)\times I_\text{out}(t).
    \end{equation}
    \item Maintain the previous value until the next time stamp (i.e.\ zero-order hold).
    \item Determine the energy consumed in one hour by integrating the power over hourly intervals, $\Delta t = 1$ hour,
    \begin{equation}
        E(t)=\int_{t-\Delta t}^{t}P(\tau) d\tau .
    \end{equation}
    \item For each household, construct a vector of hourly energy consumption values, $E$, where each element is the energy consumed in that hour ($t_{end}$ represents the final time step in the time series data), i.e.,
    \begin{equation}
        E_k = E(k \cdot \Delta t) \quad \text{where} \quad k \in \left[1, \frac{t_{\text{end}}}{\Delta t} \right].
    \end{equation}
\end{enumerate}

For the 1,000 systems considered, there are a total of 647,021 daily load profiles. Since many of these are very similar, we downsampled a subset from the total set to improve computational efficiency. Stratified sampling was used, dividing the dataset into subgroups based on daily energy consumption. First, the distribution of daily energy consumption across the entire dataset was found, then time series were randomly selected so as to keep the same distribution of daily energy consumption in the sampled subset. The process was applied twice: Initially, for each household, ten representative days were selected, giving 10,000 time series in total. This ensures that  every household and the overall spread of energy consumption are represented. Then, we applied the same method to reduce this set down to 2,000 time series in total. These steps together ensure both a representatively diverse dataset and the option to apply an algorithm to find a global optimum in the clustering process (see next section).

After the cluster centres were found, each of the 647,021 daily load profiles were assigned to their nearest cluster according to DTW distance. The end result is that each household can then represented by a chronological sequence of days, where each day is labelled with the cluster that is the closest match to the load profile of that day.

The BBOXX dataset also contains customer payment data that has previously been analysed extensively \cite{Mergulhao2023HowKenya}. For this work, the relevant payment information is the time series of remaining credit for each SHS. The contracts vary both by location and load---household energy limits are restricted based on the number of appliances owned, which relates to the customer tariff. To allow comparison between households, we convert the credit in local currency to the number of days of electricity use remaining. This removes the need for conversion between currencies and payment amounts. Once a household runs out of credit a system enters a  `disabled' state where discharging is not permitted until further payment---this is deemed an economic outage \cite{Ferrall2022MeasuringSolutions}. The clustering exercise previously described ignored this condition (i.e., clustered systems may or may not be in this stage), but we later consider it when analysing the results.

We also calculated another payment metric---the utilisation rate. This is defined as the number of days a household is not in the economic outage state divided by the number of days the household has owned the system. For example, a household with a utilisation rate of 0.9 experienced economic outages on 10\% of total days since obtaining the SHS. 

\section{Clustering Method}
\label{method}

This work uses our DTW-MIP clustering algorithm, fully described in Kumtepeli et al.\ \cite{Kumtepeli2023FastC++}. A brief overview of the approach is given here for readability and completeness.

The DTW distance (cost) between individual elements of two time series $x$ and $y$ of length $n$ and $m$ (indexed by $i$ and $j$), respectively, is given by
\begin{equation}
    \label{c}
    c_{i,j} = (x_i-y_j)^2+\min\begin{cases}
    c_{i-1,j-1}\\
    c_{i-1,j}\\
    c_{i,j-1}
    \end{cases}
\end{equation}
where the final element $c_{n,m}$ is the total DTW distance between series $x$ and $y$. This problem can be solved using dynamic programming. Here we chose not to use a warping window to restrict the parts of one time series that can be mapped to another \cite{Sakoe1978DynamicRecognition}. This allows long-range similarities to be found between peaks in energy consumption, i.e.,  whether they occur at night or in the morning. Dynamic time warping already penalises warping because more warping means more pairwise comparisons that are embedded in the final cost. Therefore the hard constraint of a window is not necessary.

The DTW distance between each time series is stored in a symmetric matrix $D_{p \times p}$, where $p$ is the total number of time series. A binary square matrix $A_{p \times p}$ is used for MIP cluster assignment, where $A_{ij} = 1$ if time series $j$ is in cluster with centroid $i$. There are $k$ clusters (\ref{eq:constraint1}), and each time series must be in one cluster only (\ref{eq:constraint2}). Clusters can be found by solving the optimisation problem 
\begin{subequations}
\begin{align}
    \min_{A} &\sum_i \sum_j D_{ij} \times A_{ij}& \\ 
    \text{subject to} \quad
     \sum_{i=1}^p A_{ii} &= k, & \label{eq:constraint1}\\
    \sum_{i=1}^p A_{ij} &= 1  &\forall j \in [1,p],  \label{eq:constraint2}\\
    A_{ij} &\leq A_{ii} &\forall i,j \in [1,p]. \label{eq:constraint3}
\end{align}
\end{subequations}

To choose the value of $k$, the algorithm was repeated for 2 to 8 clusters. We then compared the costs and silhouette scores \cite{Shahapure2020ClusterScore} to select a final value $k=5$. 

\section{Results and Discussion}

The results of this study are divided into three sections. The first presents representative daily load demand profiles identified through clustering. The second focuses on long-term changes in household energy use, first exploring consumption trends and then long-term trends in cluster allocation over the entire dataset. Finally, we consider links between payment data and electricity use.

\subsection{Daily clustering}

\begin{figure*}
    \centering
    \includegraphics[width=1\linewidth]{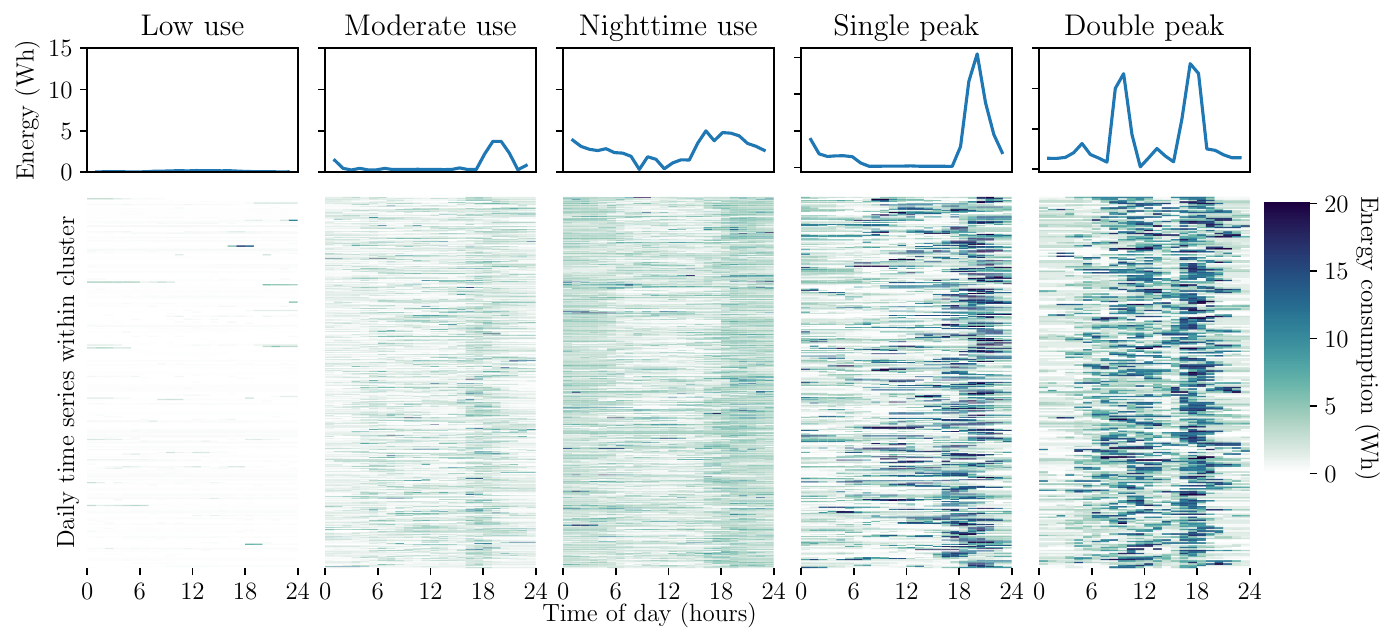}
    \caption{The 5 representative cluster centres from the DTW-MIP clustering of daily SHS load demand profiles, with the heatmaps for all time series within each cluster below. Each heatmap row represents a 24-hour load demand within that cluster. The clusters can be archetypally named: \textit{low use}, \textit{moderate use}, \textit{nighttime use}, \textit{single peak} and \textit{double peak}.}
    \label{fig:centres}
\end{figure*}
\begin{table}[]
\centering
\resizebox{\columnwidth}{!}{%
\begin{tabular}{l|lll|lll}
\hline
\multicolumn{1}{c|}{\multirow{3}{*}{Cluster}} & \multicolumn{3}{c|}{Clustering subset} & \multicolumn{3}{c}{Full dataset} \\
\multicolumn{1}{c|}{} & Count & Proportion & Mean & Count & Proportion & Mean \\ 
\multicolumn{1}{c|}{} &  & of subset & (Wh) &  & of dataset & (Wh) \\ \hline
Economic outage & - & - & - & 118258 & 0.18 & 0 \\
Low use & 259 & 0.13 & 4 & 39040 & 0.06 & 12 \\
Moderate use & 583 & 0.29 & 42 & 154765 & 0.24 & 39 \\
Nighttime use & 573 & 0.29 & 69 & 181696 & 0.28 & 73 \\
Single peak & 339 & 0.17 & 102 & 86362 & 0.14 & 102 \\
Double peak & 246 & 0.12 & 110 & 66900 & 0.10 & 122 \\
\hline
\end{tabular}%
}
\caption {Comparison of clustering results from subset vs.\ extrapolation to full dataset.}
\label{tab:c_size}
\end{table}

\begin{table}%[h!]
\resizebox{\columnwidth}{!}{%
\begin{tabular}{l|llllll}
\hline
\multicolumn{1}{c|}{\multirow{2}{*}{Cluster}} & Daily & TV own- & Weekday  & Daytime  & Appliance \\ 
        & peaks       & ership (\%)    & use (\%)        & use (\%)       & power (W)    \\ \hline
Economic outage  & 0     & 49             & 72          & N/A          & 73\\
Low use          & 0     & 48             & 72          & 50         & 71\\
Moderate use     & 0.2   & 17             & 72          & 50           & 50\\
Nighttime use    & 0.1   & 22             & 71          & 41            & 55\\
Single peak      & 1.1   & 84             & 71          & 46           & 96\\
Double peak      & 1.9   & 92             & 71          & 59            & 97\\
Total population & 0.5   & 43             & 71          & 48          & 68\\ \hline
\end{tabular}%
}
\caption{Mean characteristics of all clustered time series.}
\label{tab:c_prop}
\end{table}

Fig.\ \ref{fig:centres} depicts the centroids of the five clusters resulting from the DTW-MIP clustering exercise, and heatmaps for each cluster. The centroids are the individual time series at the median of each cluster. 
It is also helpful to examine all the time series within each cluster to gain further insight into discrepancies between and within clusters. The heatmaps present each time series in each cluster as a row, with lighter blues representing low energy consumption and darker blues indicating high consumption. Overall, the results confirm the general agreement between the cluster representatives and their associated time series.

Table \ref{tab:c_size} shows the allocation of time series across clusters within the subset, along with the results from the extrapolation to the whole dataset. The validity of the extrapolation is generally acceptable, with a larger proportion of time series in the \textit{economic outage} subset of \textit{low use}, but there is a reasonably consistency between the full dataset and subset.

Table \ref{tab:c_prop} shows characteristics of the time series within each cluster, analysed on the whole dataset, including the peaks and information on appliances and time of energy use. A peak within a time series is defined as having double the amplitude of the mean energy consumption of the time series, at least two hours apart from another distinct peak (to reduce the risk of fluctuations across the threshold counting as two peaks). Additionally, to prevent users with low consumption returning many peaks, a peak must have a minimum amplitude of \SI{10}{W}. 

The \textit{low use} cluster stands out as distinct from the other clusters, and signals that either there is an economic outage, or there is virtually no usage despite payment. To differentiate between these two scenarios, a new subset was created, classifying \textit{low use} time series as \textit{economic outage} if the credit balance was also zero. It is notable that 24\% of days in the dataset belong to \textit{economic outage} and \textit{low use} combined. Of these, 75\% (i.e., 18\%/24\%) correspond to economic outages, but there is still a significant proportion of days where the systems are paid for, but not used.

The \textit{moderate use} cluster is large and has the next lowest mean daily consumption. Typical moderate use (heatmap, Fig.\ \ref{fig:centres}) is characterised by consistent low use throughout the day, with more energy consumed in the evening and a slight increase in morning use. The evening use spans between 15:00 and 21:00 and therefore crosses sunset at around 18:00, but drops to almost no consumption after midnight. The peak demand is fairly low and few households with days in this cluster have TVs, which is reflected in the low mean appliance power. Alternatively, households with TVs and higher power appliances are not using all them on days when they are in this cluster.

\textit{Nighttime use} is another large cluster, with a slightly larger mean daily consumption than \textit{moderate use}, however there is significant overlap in the daily energy consumption between these two clusters. For days in \textit{nighttime use} the proportion of households owning TVs is low, much like for \textit{moderate use}, and there is a low appliance power of \SI{55}{W}. The characteristic that distinguishes a time series from being in \textit{moderate use} vs.\ \textit{nighttime use}, is that \textit{nighttime use} profiles consume most of their energy overnight between 18:00 and 06:00, whereas in the moderate use case electricity is consumed in the late afternoon and evening.

The final two clusters, \textit{double peak} and \textit{single peak}, represent higher consumption (households are likely to have a TV), with the double peak case slightly higher than the single. And, of course, the number of peaks is different between the two clusters---as seen in Table \ref{tab:c_prop} and Fig.\ \ref{fig:centres}. A final distinction between these two higher consumption clusters is that there is a higher proportion of daytime use in \textit{double peak}, caused by the additional morning peak alongside the evening peak. 

Finally, we include the proportion of days in each cluster that are weekdays (Table \ref{tab:c_prop}), showing that there is no significant difference between clusters---in all cases the weekday proportion is around 72\%, i.e., 5/7. This differs from HICs where there is often a noticeable difference between weekday and weekend consumption \cite{Trotta2020AnWhen}.

In summary, there is significant heterogeneity in household energy use despite the identical technical capabilities of each system (e.g., same sized battery and solar panel). The causes of this might be financial or due to the diverse energy needs of different households on different days, caused by diversity in activities. There could be a case for supplying different intensity users with differently sized systems.

\subsection{Long-term use}

\subsubsection{General consumption trend}

\begin{figure}
    \centering
    \includegraphics[width=1\linewidth]{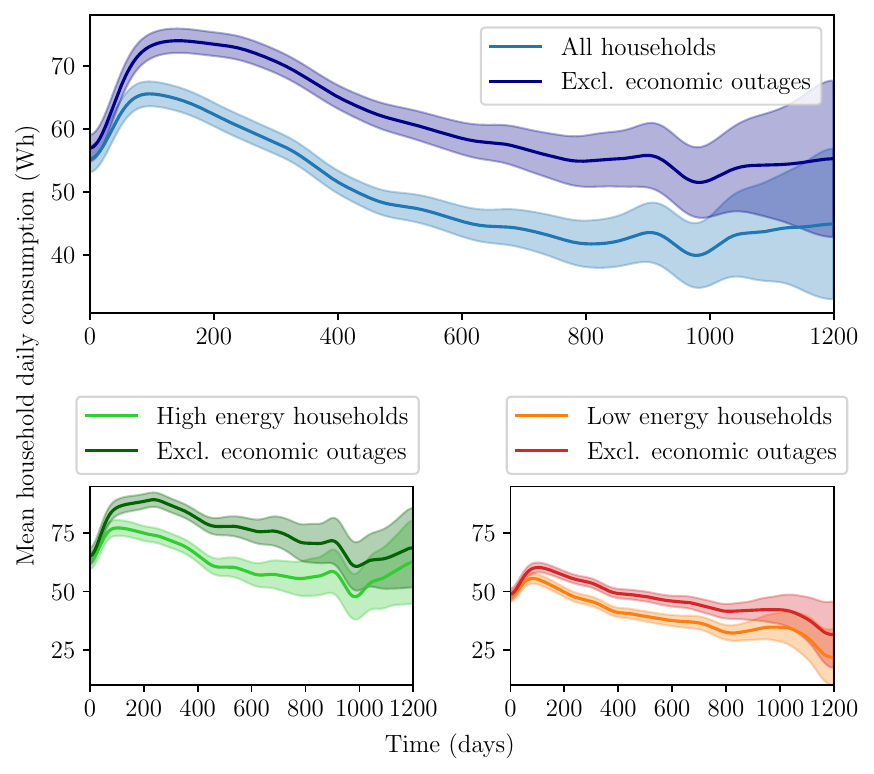}
    \caption{Top: Mean daily household electricity consumption initially increases, but then decreases. The confidence interval becomes wider over time as the number of households in the sample decreases. Bottom: High-use households ($>50$ Wh of appliances), left, vs.\ low-use ($\leq 50$ Wh of appliances), right.}
    \label{fig:use_time}
\end{figure}

Fig.\ \ref{fig:use_time} shows an initial increase in the mean daily consumption for all households, peaking on the 96th day. This is likely to be a `settling in period' where households also purchase appliances and become familiar with the system. %
Over the first six months, daily electricity consumption increases for 58\% of customers, but by the end of the first year, only 43\% of customers continue to increase consumption. Beyond the first year, 77\% of users are decreasing daily consumption. This decrease is particularly prominent in the first 400 days, and then diverges depending on household energy use. The mean energy consumption over two years shows a 33\% reduction from its peak, dropping from \SI{66}{Wh/day} to \SI{44}{Wh/day}. This is contrary to prior literature where the expectation is that consumption will continue to increase after a system is acquired, since users will obtain more appliances and find increasingly energy-intensive activities to engage with \cite{Dominguez2021UnderstandingHouseholds}. Instead of this, our results show that individual SHS users  follow a long-term consumption trend similar to that witnessed for individual grid-connected users \cite{Masselus202410Rwanda, Mugyenyi2025Post-connectionRwanda}.

Mean consumption in high-use households begins to plateau beyond 400 days. When economic outages are removed the decrease in consumption is slightly stronger, especially beyond 700 days. Another important observation is that when economic outages are ignored, the consumption in high-use households continues to increase until around 250 days, i.e., the apparent decrease in the mean consumption of high-use households after the peak around day 96 is actually caused by a high proportion of economic outages.

In low-use households, the mean relative decrease to the end of the first year is even greater (24\%) than in high-use households (12\%). The decrease continues, although at a lower rate, until around 800 days. The difference between the mean consumption with and without economic outages in low-use households grows quickly in the first part of ownership and then gradually increases over time.

A gradual decrease in electricity use can lead to oversized systems. This is undesirable because the full solar generation potential is not realised \cite{Bhatti2021EstimationSystems, Soltowski2019Bottom-upRwanda}, batteries are not cycled optimally, shortening lifetime \cite{Perriment2023Lead-AcidConditions, Narayan2019ExploringAccess}, and customers pay for a larger system than needed, increasing their electricity cost. This is particularly detrimental because the cost of energy is a significant barrier to electrification \cite{Kyere2024DecodingPhotovoltaic, Kizilcec2020SolarAfrica}.

\subsubsection{Cluster allocation trend}

We now analyse long-term trends in cluster allocation for individual households over time. An interesting finding is a lack of homogeneity within households, where homogeneity is defined as the fraction of days that a household's demand is in its dominant cluster (the cluster the household spends the most time in). Table \ref{tab:dom_c} shows that the mean homogeneity for all households is only 55\%.

\begin{table}%[h!]
\resizebox{\columnwidth}{!}{%
\begin{tabular}{l|lllll}
\hline
Dominant cluster & Homogen- & Utilisation& TV own-  & Flexible & Appliance \\ 
        &   eity (\%) & rate (\%) & ership (\%)     & appliances & power (W)    \\ \hline
Economic outage  & 51   &48  & 62  & 1.4  & 82       \\
Low use          & 42   &84  & 53  & 1.1 & 72         \\
Moderate use     & 56 &89  & 8    & 0.7  & 44         \\
Nighttime use    & 61  &90 & 19  & 0.7  & 57        \\
Single peak      & 46 &93 & 91  & 1.9  & 102      \\
Double peak      & 48 &92  & 98   & 1.8  & 99      \\
\hline
\end{tabular}%
}
\caption{Household mean characteristics separated by dominant cluster---where they spend the majority of time.}
\label{tab:dom_c}
\end{table}

Higher consumption clusters (\textit{single peak} and \textit{double peak}) exhibit lower homogeneity. Table \ref{tab:dom_c} also shows that households with days in higher consumption clusters have greater appliance power, with the majority owning a TV. Therefore, households with days in higher consumption dominant clusters have the potential to be in all clusters, however, the majority of households with other dominant clusters do not consume larger amounts of energy due to lack of appliances. Additionally, in the higher consumption clusters, households have more flexible loads. Base households have lights and phone chargers and these are relatively inflexible because lighting is needed when it is dark and phone charging when batteries are depleted. Here, we define anything other than lighting and phone charging to be a flexible load. In the higher consumption clusters, households have appliances like TVs and radios which they may not use every day, leading to more frequent day-to-day changes in electricity consumption.

\begin{figure}
    \centering
    \includegraphics[width=0.95\linewidth]{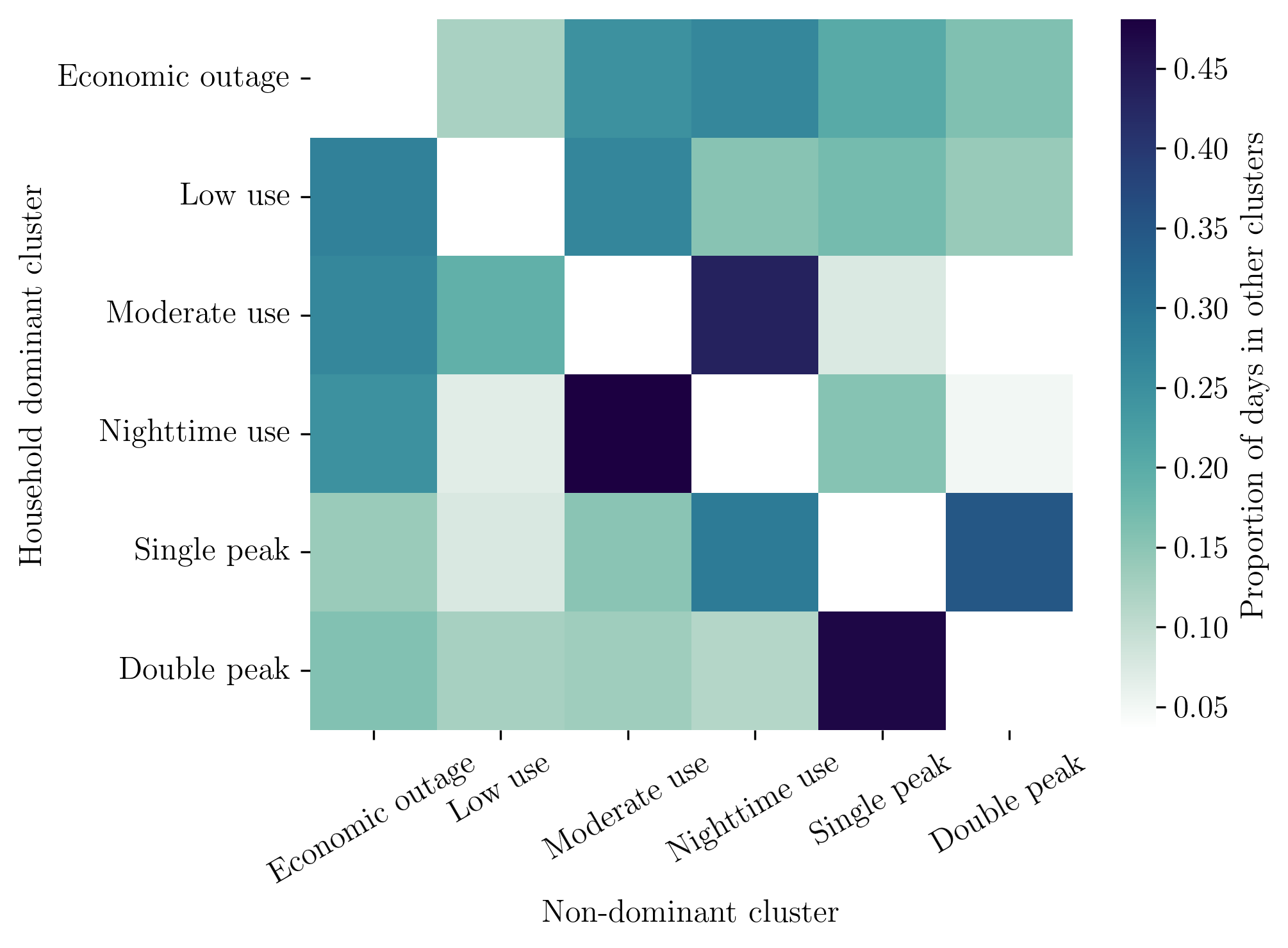}
    \caption{Spread of usage over clusters: for each household with a certain dominant cluster (rows), the proportion of time spent in other clusters is indicated by the columns.}
    \label{fig:common_clus}
\end{figure}

Fig.\ \ref{fig:common_clus} depicts how households, split into their dominant clusters, consume energy when not following the dominant cluster behaviour. This further demonstrates that households dominated by higher consumption can also have usage days in all other clusters, whereas is it very uncommon for \textit{moderate use} dominated households to have days in the \textit{single peak} and \textit{double peak} clusters.

\textit{Economic outage} is observed in all households, but is particularly prevalent in the three lower consumption dominated cases (\textit{low use}, \textit{moderate use} and \textit{nighttime use}). This is logical, because customers who initially pay less have access to smaller amounts of available energy, and fewer appliances, and may be more likely to face financial hardship or be less engaged with their system---both of which lead to missed payments and economic outages.

Fig.\ \ref{fig:common_clus} also shows that clusters with similar average consumption are more closely linked within households (e.g., \textit{moderate use} and \textit{nighttime use}, similarly \textit{single peak} and \textit{double peak}). Therefore, while household energy consumption is clearly non-uniform, it is more common for households to vary their consumption behaviour within similar magnitudes of use or to not use the system at all and not pay.

\begin{figure}
    \centering
    \includegraphics[width=1\linewidth]{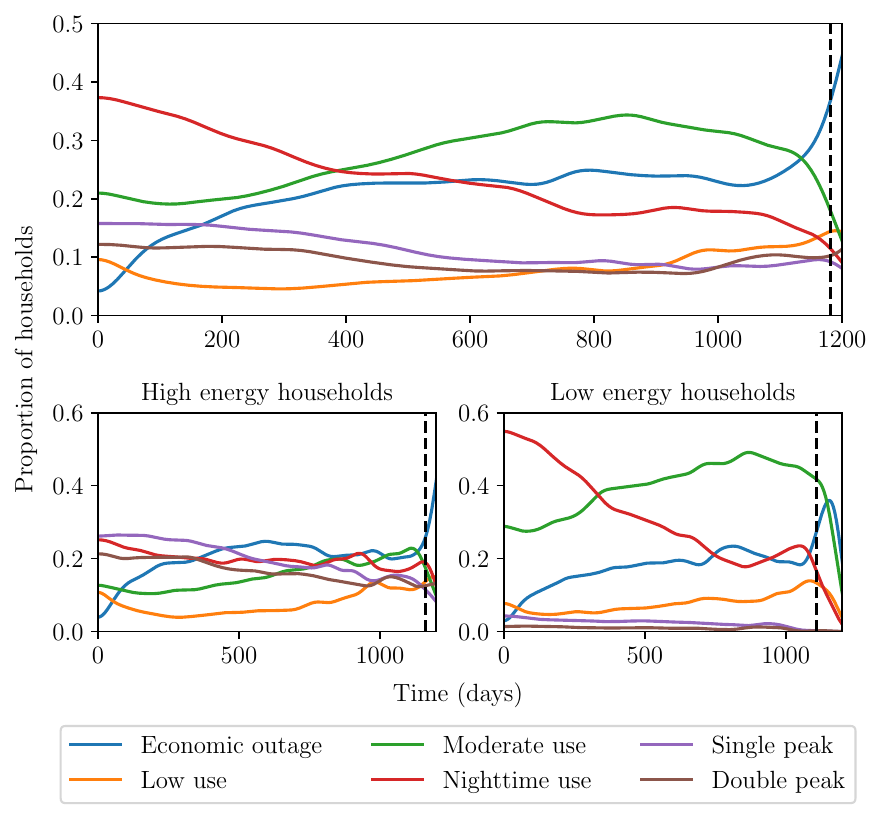}
    \caption{Top: Proportion of households in each cluster, every day. Bottom: High-use households ($>50$ Wh of appliances), left, and low-use ($\leq 50$ Wh of appliances), right. Black dashed line indicates when number of households drops below 10.}
    \label{fig:cluster_time}
\end{figure}

Fig.\ \ref{fig:cluster_time} shows the proportion of households in each cluster on each day. The most important finding is the sharp increase in \textit{economic outage} over the first year. This is the primary explanation reduced electricity consumption, as previously seen in Fig.\ \ref{fig:use_time}, and shows that customers either cannot afford to keep up with payments or are choosing not to spend their money on SHS credit. Once \textit{economic outage} begins to plateau, \textit{low use} starts to gradually increase. The \textit{low use} cluster shows when households are choosing to use the system minimally regardless of economic considerations. The increase therefore shows disengagement with the system for non-economic reasons. 

Previously, Fig.\ \ref{fig:common_clus}, a link between \textit{nighttime use} and \textit{moderate use} was observed. This is reinforced in Fig.\ \ref{fig:cluster_time}, especially in the low-use subplot, where \textit{nighttime use} significantly reduces over the time of analysis, while \textit{moderate use} increases. We hypothesise that this is the driver for the decreasing use in low-use households shown in Fig.\ \ref{fig:use_time}. Given their already low consumption, a shift from \textit{nighttime use} to \textit{moderate use} has a substantial impact. For instance, households might initially use lighting throughout the night for security purposes, but over time, reduce this to evening hours only. To fully understand the reasons behind this behavioural change, further research including surveys is necessary.

The high energy household subplot in Fig.\ \ref{fig:cluster_time} shows a similar trend of consistently decreasing \textit{single peak} and \textit{double peak} with increase in \textit{moderate use} and \textit{low use}. Thus, households reduce peak consumption over time and reduce the use of flexible loads.

\subsection{Utilisation rate and electricity use}

\begin{figure}
    \centering
    \includegraphics[width=1\linewidth]{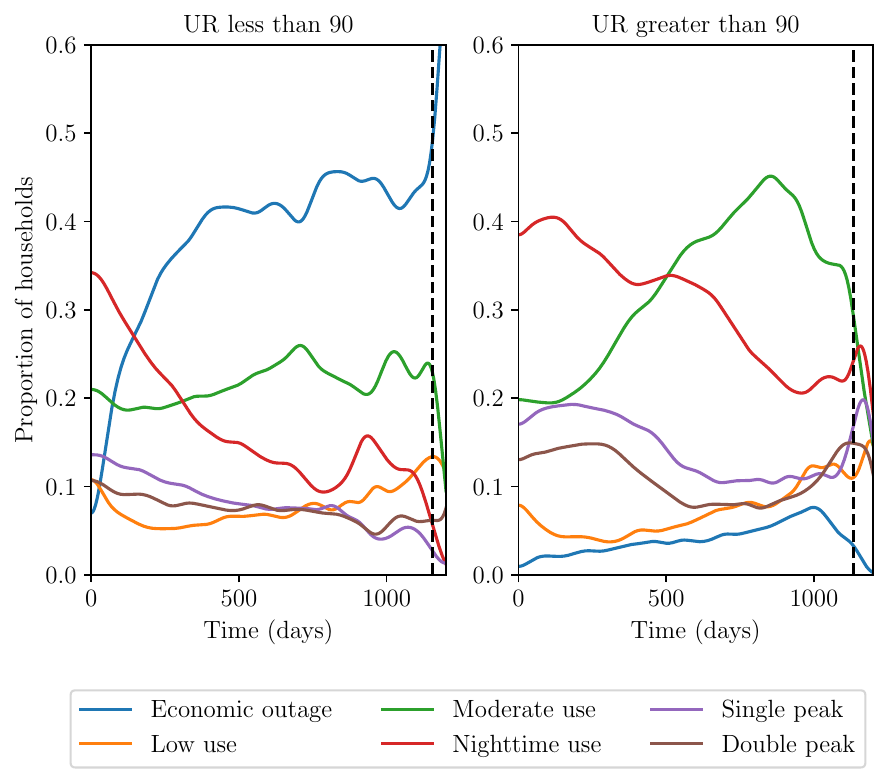}
    \caption{Proportion of households in each cluster over time, split into households with a utilisation rate over 90\% and under 90\% at time of analysis. The black dashed line indicates when the number of households in the sample goes below 10.}
    \label{fig:ur_clus}
\end{figure}

Fig.\ \ref{fig:cluster_time} illustrates a rise in \textit{economic outage} over time. However, a deeper understanding of household behaviour can be gained by examining utilisation rates. Fig.\ \ref{fig:ur_clus} presents the time variation in the proportion of households within each cluster segmented by utilisation rate. There are 530 households in the dataset with a utilisation rate of 0.9 and above, but only 121 households with a utilisation rate of less than 0.5. 

The observed increase in \textit{economic outage} is primarily driven by households with lower utilisation rates, which, by definition, is less evident in households with higher utilisation rates. In the low utilisation rate group, \textit{nighttime use} significantly reduces, particularly in the first year; \textit{low use} and \textit{moderate use} fluctuate but stay relatively constant, whereas \textit{single peak} and \textit{double peak} steadily decrease. In low utilisation households, there isn't a significant increase in any cluster aside from \textit{economic outage}, so these households are clearly changing their energy behaviour for  financial reasons---either an affordability issue or a lack of desire to spend money on the system.

For households with higher utilisation rate, the proportion of \textit{nighttime use} continually decreases but at a steadier rate than in lower utilisation households. Higher utilisation rate households are more likely to be in \textit{single peak} and \textit{double peak} clusters than lower utilisation rate households. The mean proportion of households in these two high-consumption clusters across the whole time period is 19\% for low utilisation rate households and 32\% for high utilisation households. Thus, households that pay more consistently are more likely to be in the higher use clusters. However, beyond the first year, \textit{single peak} and \textit{double peak} rates decrease, while \textit{low use} and \textit{moderate use} increase. This demonstrates how households are reducing their electricity consumption even when \textit{economic outage} is negligible, which further builds on the previously stated hypothesis that households decrease electricity consumption even while still paying for and engaging with the SHS.

Overall, splitting the households by utilisation rates highlights several key issues: 
\begin{itemize}
    \item Energy consumption patterns vary with utilisation rate.
    \item \textit{Economic outage} increases significantly over the first year of ownership in low utilisation households. In high utilisation households, the small proportion of  outages increases gradually over time, peaking at the 3-year mark.
    \item There is a shift to lower consumption clusters over time and it is more pronounced in households with utilisation rate over 90\%.
    \item Higher consumption clusters are more common in higher utilisation rate households.
\end{itemize}

\section{Conclusions}

This work has analysed a large dataset of energy use in 1,000 solar home systems across sub-Saharan Africa. Time series clustering identified five distinct daily usage profiles, archetypally defined as \textit{low use}, \textit{moderate use}, \textit{nighttime use}, \textit{single peak} and \textit{double peak}, in order of increasing mean energy use. The representation of these across the dataset shows that systems are only fully utilised, i.e., their energy usage is in the \textit{single peak} or \textit{double peak} cluster, on 24\% of days---the same proportion of days that the systems are in the \textit{economic outage} and \textit{low use} clusters. The majority of time, the systems provide basic supply of \textit{moderate use} and \textit{nighttime use} electricity. Households frequently change, from day to day, how they use their systems, with cluster homogeneity for households averaging only 55\%.

Analysis of trends in household behaviour over a period of up to 1,200 days revealed a general decrease in energy consumption among both high and low consumption households. While economic outages are not uncommon under the PAYG scheme, they are not the sole reason for decreasing energy consumption, indicating that changes in energy behaviour extend beyond financial considerations. The decrease in energy consumption when not in the economic outage state is characterised by a decrease in nighttime energy use, and reduction of days with large evening peaks. Utilisation rates are slightly higher for households that predominantly exhibit \textit{single peak} and \textit{double peak} profiles, but even these households experience economic outages.

Overall, the systems are often underused and therefore oversized, leading to knock-on effects on costs for households and suppliers. The significant proportion of economic outages suggests that costs remain a limitation for many households, warranting further investigation into how to maximise energy provision from each SHS in order to reduce the overall cost of energy. There is a need for SHS designs and pricing structures to better align with the actual energy requirements of households. Adaptable solutions may be key for long term adoption, and this may include downsizing systems over time in some cases.

Further work should include surveys of households to understand the underlying causes of the non-financial decrease in energy consumption. Moreover, continuous study of energy behaviour as households gain access to electricity is essential for better understanding their needs and how these evolve over time.

\section{Acknowledgements}
The authors thank BBOXX for project funding and access to SHS data, and EPSRC for project funding (ref.\ EP/W027321/1). Prof.\ Parikh is grateful to the Royal Academy of Engineering for part funding.

For the purpose of Open Access, the authors apply a CC BY public copyright licence to any Author Accepted Manuscript version arising from this submission.

\small
\bibliography{references}

\begin{thebibliography}{10}
\providecommand{\url}[1]{#1}
\csname url@samestyle\endcsname
\providecommand{\newblock}{\relax}
\providecommand{\bibinfo}[2]{#2}
\providecommand{\BIBentrySTDinterwordspacing}{\spaceskip=0pt\relax}
\providecommand{\BIBentryALTinterwordstretchfactor}{4}
\providecommand{\BIBentryALTinterwordspacing}{\spaceskip=\fontdimen2\font plus
\BIBentryALTinterwordstretchfactor\fontdimen3\font minus \fontdimen4\font\relax}
\providecommand{\BIBforeignlanguage}[2]{{%
\expandafter\ifx\csname l@#1\endcsname\relax
\typeout{** WARNING: IEEEtran.bst: No hyphenation pattern has been}%
\typeout{** loaded for the language `#1'. Using the pattern for}%
\typeout{** the default language instead.}%
\else
\language=\csname l@#1\endcsname
\fi
#2}}
\providecommand{\BIBdecl}{\relax}
\BIBdecl

\bibitem{GoalAffairs}
\BIBentryALTinterwordspacing
``{Goal 7 Department of Economic and Social Affairs}.'' [Online]. Available: \url{https://sdgs.un.org/goals/goal7}
\BIBentrySTDinterwordspacing

\bibitem{Almeshqab2019LessonsAspects}
F.~Almeshqab and T.~S. Ustun, ``{Lessons learned from rural electrification initiatives in developing countries: Insights for technical, social, financial and public policy aspects},'' pp. 35--53, 3 2019.

\bibitem{Narayan2020TheChallenges}
N.~Narayan, V.~Vega-Garita, Z.~Qin, J.~Popovic-Gerber, P.~Bauer, and M.~Zeman, ``{The long road to universal electrification: A critical look at present pathways and challenges},'' \emph{Energies}, vol.~13, no.~3, 2020.

\bibitem{Irena2022RENEWABLEREGIONS}
\BIBentryALTinterwordspacing
{Irena}, \emph{{RENEWABLE ENERGY MARKET ANALYSIS AFRICA AND ITS REGIONS}}, 2022. [Online]. Available: \url{www.afdb.org}
\BIBentrySTDinterwordspacing

\bibitem{Pandyaswargo2020EstimatingLaos}
A.~H. Pandyaswargo, M.~Ruan, E.~Htwe, M.~Hiratsuka, A.~D. Wibowo, Y.~Nagai, and H.~Onoda, ``{Estimating the energy demand and growth in off-grid villages: Case studies from Myanmar, Indonesia, and Laos},'' \emph{Energies}, vol.~13, no.~20, 10 2020.

\bibitem{Sandwell2016AnalysisPradesh}
P.~Sandwell, C.~Chambon, A.~Saraogi, A.~Chabenat, M.~Mazur, N.~Ekins-Daukes, and J.~Nelson, ``{Analysis of energy access and impact of modern energy sources in unelectrified villages in Uttar Pradesh},'' \emph{Energy for Sustainable Development}, vol.~35, pp. 67--79, 12 2016.

\bibitem{Mandelli2016NovelAreas}
S.~Mandelli, M.~Merlo, and E.~Colombo, ``{Novel procedure to formulate load profiles for off-grid rural areas},'' \emph{Energy for Sustainable Development}, vol.~31, pp. 130--142, 4 2016.

\bibitem{Boait2017ESCoBox:World}
P.~Boait, R.~Gammon, V.~Advani, N.~Wade, D.~Greenwood, and P.~Davison, ``{ESCoBox: A set of tools for mini-grid sustainability in the developing world},'' \emph{Sustainability (Switzerland)}, vol.~9, no.~5, 2017.

\bibitem{Boait2015EstimationCountries}
P.~Boait, V.~Advani, and R.~Gammon, ``{Estimation of demand diversity and daily demand profile for off-grid electrification in developing countries},'' \emph{Energy for Sustainable Development}, vol.~29, pp. 135--141, 12 2015.

\bibitem{Blodgett2017AccuracyUserconsumption}
C.~Blodgett, P.~Dauenhauer, H.~Louie, and L.~Kickham, ``{Accuracy of energy-use surveys in predicting rural mini-grid user consumption},'' \emph{Energy for Sustainable Development}, vol.~41, pp. 88--105, 12 2017.

\bibitem{Hartvigsson2018ComparisonData}
E.~Hartvigsson and E.~O. Ahlgren, ``{Comparison of load profiles in a mini-grid: Assessment of performance metrics using measured and interview-based data},'' \emph{Energy for Sustainable Development}, vol.~43, pp. 186--195, 4 2018.

\bibitem{Allee2021PredictingModels}
A.~Allee, N.~J. Williams, A.~Davis, and P.~Jaramillo, ``{Predicting initial electricity demand in off-grid Tanzanian communities using customer survey data and machine learning models},'' \emph{Energy for Sustainable Development}, vol.~62, pp. 56--66, 6 2021.

\bibitem{Gelchu2023ComparisonEthiopia}
M.~A. Gelchu, J.~Ehnberg, and E.~O. Ahlgren, ``{Comparison Of Electricity Load Estimation Methods In Rural Mini-Grids: Case Study In Ethiopia}.''\hskip 1em plus 0.5em minus 0.4em\relax Institute of Electrical and Electronics Engineers (IEEE), 12 2023, pp. 1--5.

\bibitem{Bisaga2017ScalableSolar}
I.~Bisaga, N.~Pu{\'{z}}niak-Holford, A.~Grealish, C.~Baker-Brian, and P.~Parikh, ``{Scalable off-grid energy services enabled by IoT: A case study of BBOXX SMART Solar},'' \emph{Energy Policy}, vol. 109, pp. 199--207, 2017.

\bibitem{Few2022ElectricityDesign}
S.~Few, J.~Barton, P.~Sandwell, R.~Mori, P.~Kulkarni, M.~Thomson, J.~Nelson, and C.~Candelise, ``{Electricity demand in populations gaining access: Impact of rurality and climatic conditions, and implications for microgrid design},'' \emph{Energy for Sustainable Development}, vol.~66, pp. 151--164, 2 2022.

\bibitem{Bhattacharyya2021AAsia}
S.~C. Bhattacharyya and D.~Palit, ``{A critical review of literature on the nexus between central grid and off-grid solutions for expanding access to electricity in Sub-Saharan Africa and South Asia},'' 5 2021.

\bibitem{DenHeeten2017UnderstandingElectrification}
T.~Den~Heeten, N.~Narayan, J.~C. Diehl, J.~Verschelling, S.~Silvester, J.~Popovic-Gerber, P.~Bauer, and M.~Zeman, ``{Understanding the present and the future electricity needs: Consequences for design of future Solar Home Systems for off-grid rural electrification},'' in \emph{Proceedings of the 25th Conference on the Domestic Use of Energy, DUE 2017}.\hskip 1em plus 0.5em minus 0.4em\relax Institute of Electrical and Electronics Engineers Inc., 5 2017, pp. 8--15.

\bibitem{Jurasz2022OnSystems}
J.~Jurasz, M.~Guezgouz, P.~E. Campana, and A.~Kies, ``{On the impact of load profile data on the optimization results of off-grid energy systems},'' \emph{Renewable and Sustainable Energy Reviews}, vol. 159, 5 2022.

\bibitem{Yilmaz2019ComparisonManagement}
S.~Yilmaz, J.~Chambers, and M.~K. Patel, ``{Comparison of clustering approaches for domestic electricity load profile characterisation - Implications for demand side management},'' \emph{Energy}, vol. 180, pp. 665--677, 8 2019.

\bibitem{Rajabi2020ASegmentation}
A.~Rajabi, M.~Eskandari, M.~J. Ghadi, L.~Li, J.~Zhang, and P.~Siano, ``{A comparative study of clustering techniques for electrical load pattern segmentation},'' \emph{Renewable and Sustainable Energy Reviews}, vol. 120, 3 2020.

\bibitem{Wang2015LoadReview}
Y.~Wang, Q.~Chen, C.~Kang, M.~Zhang, K.~Wang, and Y.~Zhao, ``{Load Profiling and Its Application to Demand Response: A Review},'' Tech. Rep.~2, 2015.

\bibitem{McLoughlin2015AData}
F.~McLoughlin, A.~Duffy, and M.~Conlon, ``{A clustering approach to domestic electricity load profile characterisation using smart metering data},'' \emph{Applied Energy}, vol. 141, pp. 190--199, 3 2015.

\bibitem{VonLuxburg2012Clustering:Art}
U.~Von~Luxburg and R.~C. Williamson, ``{Clustering: Science or Art?}'' Tech. Rep., 2012.

\bibitem{Hennig2015WhatClusters}
C.~Hennig, ``{What are the true clusters?}'' \emph{Pattern Recognition Letters}, vol.~64, pp. 53--62, 10 2015.

\bibitem{Sakoe1978DynamicRecognition}
H.~Sakoe and S.~Chiba, ``{Dynamic Programming Algorithm Optimization for Spoken Word Recognition},'' \emph{IEEE Transactions on Acoustics, Speech, and Signal Processing}, vol.~26, no.~1, pp. 43--49, 1978.

\bibitem{Lin2019ClusteringApplications}
S.~Lin, F.~Li, E.~Tian, Y.~Fu, and D.~Li, ``{Clustering load profiles for demand response applications},'' \emph{IEEE Transactions on Smart Grid}, vol.~10, no.~2, pp. 1599--1607, 3 2019.

\bibitem{Aghabozorgi2015Time-seriesReview}
S.~Aghabozorgi, A.~Seyed~Shirkhorshidi, and T.~Ying~Wah, ``{Time-series clustering - A decade review},'' \emph{Information Systems}, vol.~53, pp. 16--38, 5 2015.

\bibitem{Ausmus2020ImprovingWarping}
J.~R. Ausmus, P.~K. Sen, T.~Wu, U.~Adhikari, Y.~Zhang, and V.~Krishnan, ``{Improving the Accuracy of Clustering Electric Utility Net Load Data using Dynamic Time Warping},'' in \emph{Proceedings of the IEEE Power Engineering Society Transmission and Distribution Conference}, vol. 2020-October.\hskip 1em plus 0.5em minus 0.4em\relax Institute of Electrical and Electronics Engineers Inc., 10 2020.

\bibitem{Begum2015AcceleratingStrategy}
N.~Begum, L.~Ulanova, J.~Wang, and E.~Keogh, ``{Accelerating dynamic time warping clustering with a novel admissible pruning strategy},'' in \emph{Proceedings of the ACM SIGKDD International Conference on Knowledge Discovery and Data Mining}, vol. 2015-August.\hskip 1em plus 0.5em minus 0.4em\relax Association for Computing Machinery, 8 2015, pp. 49--58.

\bibitem{Ratanamahatana2005ThreeMining}
\BIBentryALTinterwordspacing
C.~A. Ratanamahatana and E.~Keogh, ``{Three Myths about Dynamic Time Warping Data Mining}.''\hskip 1em plus 0.5em minus 0.4em\relax Proceedings of the 2005 SIAM International Conference on Data Mining (SDM), 2005. [Online]. Available: \url{https://epubs.siam.org/terms-privacy}
\BIBentrySTDinterwordspacing

\bibitem{Teeraratkul2016CondensedDemand}
T.~Teeraratkul, D.~O'Neilly, and S.~Lallz, ``{Condensed representation and individual prediction of consumer demand},'' in \emph{2016 4th IEEE International Conference on Smart Energy Grid Engineering, SEGE 2016}.\hskip 1em plus 0.5em minus 0.4em\relax Institute of Electrical and Electronics Engineers Inc., 10 2016, pp. 11--16.

\bibitem{Kumtepeli2024DTW-C++:Data}
\BIBentryALTinterwordspacing
V.~Kumtepeli, R.~Perriment, and D.~A. Howey, ``{DTW-C++: Fast dynamic time warping and clustering of time series data},'' \emph{Journal of Open Source Software}, vol.~9, no. 101, p. 6881, 9 2024. [Online]. Available: \url{https://joss.theoj.org/papers/10.21105/joss.06881}
\BIBentrySTDinterwordspacing

\bibitem{Williams2017LoadMicrogrids}
N.~Williams, P.~Jaramillo, B.~Cornell, I.~Lyons-Galante, and E.~Wynn, ``{Load Characteristics of East African Microgrids},'' \emph{IEEE PES-IAS PowerAfrica Confrence}, 2017.

\bibitem{Lorenzoni2020ClassificationApproach}
L.~Lorenzoni, P.~Cherubini, D.~Fioriti, D.~Poli, A.~Micangeli, and R.~Giglioli, ``{Classification and modeling of load profiles of isolated mini-grids in developing countries: A data-driven approach},'' \emph{Energy for Sustainable Development}, vol.~59, pp. 208--225, 12 2020.

\bibitem{Lukuyu2023PurchasingAfrica}
J.~Lukuyu, M.~Shiran, R.~Kennedy, J.~Urpelainen, and J.~Taneja, ``{Purchasing power: Examining customer profiles and patterns for decentralized electricity systems in East Africa},'' \emph{Energy Policy}, vol. 172, 1 2023.

\bibitem{Riva2018Long-termPerspective}
F.~Riva, A.~Tognollo, F.~Gardumi, and E.~Colombo, ``{Long-term energy planning and demand forecast in remote areas of developing countries: Classification of case studies and insights from a modelling perspective},'' pp. 71--89, 4 2018.

\bibitem{Muhumuza2018EnergyCountries}
R.~Muhumuza, A.~Zacharopoulos, J.~D. Mondol, M.~Smyth, and A.~Pugsley, ``{Energy consumption levels and technical approaches for supporting development of alternative energy technologies for rural sectors of developing countries},'' pp. 90--102, 12 2018.

\bibitem{Opiyo2020HowCommunities}
N.~N. Opiyo, ``{How basic access to electricity stimulates temporally increasing load demands by households in rural developing communities},'' \emph{Energy for Sustainable Development}, vol.~59, pp. 97--106, 12 2020.

\bibitem{Riva2019ModellingPlanning}
F.~Riva, F.~D. Sanvito, F.~Tonini, and E.~Colombo, ``{Modelling long-term electricity load demand for rural electrification planning; Modelling long-term electricity load demand for rural electrification planning},'' Tech. Rep., 2019.

\bibitem{Bisaga2018ToLens}
I.~Bisaga and P.~Parikh, ``{To climb or not to climb? Investigating energy use behaviour among Solar Home System adopters through energy ladder and social practice lens},'' \emph{Energy Research and Social Science}, vol.~44, pp. 293--303, 10 2018.

\bibitem{Kizilcec2022ForecastingAccess}
V.~Kizilcec, C.~Spataru, A.~Lipani, and P.~Parikh, ``{Forecasting Solar Home System Customers’ Electricity Usage with a 3D Convolutional Neural Network to Improve Energy Access},'' \emph{Energies}, vol.~15, no.~3, 2 2022.

\bibitem{Dominguez2021UnderstandingHouseholds}
C.~Dominguez, K.~Orehounig, and J.~Carmeliet, ``{Understanding the path towards a clean energy transition and post-electrification patterns of rural households},'' \emph{Energy for Sustainable Development}, vol.~61, pp. 46--64, 4 2021.

\bibitem{Masselus202410Rwanda}
L.~Masselus, J.~Ankel-Peters, G.~Gonzalez~Sutil, V.~Modi, J.~Mugyenyi, A.~Munyehirwe, N.~Williams, and M.~Sievert, ``{10 Years After: Long-term Adoption of Electricity in Rural Rwanda},'' Tech. Rep., 2024.

\bibitem{Louie2023DailyNation}
H.~Louie, S.~Atcitty, D.~Terry, D.~Lee, and P.~Romine, ``{Daily electrical energy consumption characteristics and design implications for off-grid homes on the Navajo Nation},'' \emph{Energy for Sustainable Development}, vol.~73, pp. 315--325, 4 2023.

\bibitem{Blimpo2020WhyAfrica}
M.~P. Blimpo, A.~Postepska, and Y.~Xu, ``{Why is household electricity uptake low in Sub-Saharan Africa?}'' \emph{World Development}, vol. 133, 9 2020.

\bibitem{Kizilcec2020SolarAfrica}
V.~Kizilcec and P.~Parikh, ``{Solar Home Systems: A comprehensive literature review for Sub-Saharan Africa},'' pp. 78--89, 10 2020.

\bibitem{Mergulhao2023HowKenya}
V.~P. Mergulh{\~{a}}o, L.~Capra, K.~Voglitsis, and P.~Parikh, ``{How do they pay as they go?: Learning payment patterns from solar home system users data in Rwanda and Kenya},'' \emph{Energy for Sustainable Development}, vol.~76, 10 2023.

\bibitem{Guajardo2019HowEconomies}
J.~A. Guajardo, ``{How Do Usage and Payment Behavior Interact in Rent-to-Own Business Models? Evidence from Developing Economies},'' \emph{Production and Operations Management}, vol.~28, no.~11, pp. 2808--2822, 11 2019.

\bibitem{Ferrall2022MeasuringSolutions}
I.~Ferrall, D.~Callaway, and D.~M. Kammen, ``{Measuring the reliability of SDG 7: The reasons, timing, and fairness of outage distribution for household electricity access solutions},'' \emph{Environmental Research Communications}, vol.~4, no.~5, 5 2022.

\bibitem{Kumtepeli2023FastC++}
\BIBentryALTinterwordspacing
V.~Kumtepeli, R.~Perriment, and D.~A. Howey, ``{Fast dynamic time warping and clustering in C++},'' 7 2023. [Online]. Available: \url{http://arxiv.org/abs/2307.04904}
\BIBentrySTDinterwordspacing

\bibitem{Shahapure2020ClusterScore}
K.~R. Shahapure and C.~Nicholas, ``{Cluster quality analysis using silhouette score},'' in \emph{Proceedings - 2020 IEEE 7th International Conference on Data Science and Advanced Analytics, DSAA 2020}.\hskip 1em plus 0.5em minus 0.4em\relax Institute of Electrical and Electronics Engineers Inc., 10 2020, pp. 747--748.

\bibitem{Trotta2020AnWhen}
G.~Trotta, ``{An empirical analysis of domestic electricity load profiles: Who consumes how much and when?}'' \emph{Applied Energy}, vol. 275, 10 2020.

\bibitem{Mugyenyi2025Post-connectionRwanda}
J.~Mugyenyi, B.~Muhwezi, S.~Fobi, C.~Massa, J.~Taneja, N.~J. Williams, and V.~Modi, ``{Post-connection electricity demand and pricing in newly electrified households: Insights from a large-scale dataset in Rwanda},'' \emph{Energy Policy}, vol. 198, 3 2025.

\bibitem{Bhatti2021EstimationSystems}
S.~Bhatti and A.~Williams, ``{Estimation of surplus energy in off-grid solar home systems},'' \emph{Renewable Energy and Environmental Sustainability}, vol.~6, p.~25, 2021.

\bibitem{Soltowski2019Bottom-upRwanda}
B.~Soltowski, D.~Campos-Gaona, S.~Strachan, and O.~Anaya-Lara, ``{Bottom-up electrification introducing new smart grids architecture-concept based on feasibility studies conducted in Rwanda},'' \emph{Energies}, vol.~12, no.~12, 2019.

\bibitem{Perriment2023Lead-AcidConditions}
R.~Perriment, V.~Kumtepeli, M.~McCulloch, and D.~Howey, ``{Lead-Acid Battery Lifetime Extension in Solar Home Systems Under Different Operating Conditions}.''\hskip 1em plus 0.5em minus 0.4em\relax Institute of Electrical and Electronics Engineers (IEEE), 12 2023, pp. 1--5.

\bibitem{Narayan2019ExploringAccess}
N.~Narayan, A.~Chamseddine, V.~Vega-Garita, Z.~Qin, J.~Popovic-Gerber, P.~Bauer, and M.~Zeman, ``{Exploring the boundaries of Solar Home Systems (SHS) for off-grid electrification: Optimal SHS sizing for the multi-tier framework for household electricity access},'' \emph{Applied Energy}, vol. 240, pp. 907--917, 4 2019.

\bibitem{Kyere2024DecodingPhotovoltaic}
F.~Kyere, S.~Dongying, G.~D. Bampoe, N.~Y.~G. Kumah, and D.~Asante, ``{Decoding the shift: Assessing household energy transition and unravelling the reasons for resistance or adoption of solar photovoltaic},'' \emph{Technological Forecasting and Social Change}, vol. 198, 1 2024.

\end{thebibliography}
\bibliographystyle{IEEEtran}

\end{document}